\documentclass[runningheads]{llncs}

\usepackage{graphicx}
\usepackage{color,soul}
\usepackage{textcomp} 
\usepackage{subfigure}

\usepackage{fancyhdr}
\begin{document}

\title{GA-Novo: \textit{De Novo} Peptide Sequencing via Tandem Mass Spectrometry using \\ Genetic Algorithm }

\author{\vspace{-1.5cm}\inst{}}

\institute{  }

\author{Samaneh Azari\inst{1}\and
Bing Xue\inst{1} \and
Mengjie~Zhang\inst{1}\and Lifeng~Peng\inst{2}}

\institute{School of Engineering and Computer Science, Victoria University of Wellington, \\ PO Box 600, Wellington 6140, New Zealand \\
	 \email{\{samaneh.azari,bing.xue,mengjie.zhang\}@ecs.vuw.ac.nz}
		 \and
		  Centre for Biodiscovery and School of Biological Sciences, Victoria University of Wellington, PO Box 600,
		  Wellington 6140, New Zealand
		 \\
		 \email{lifeng.peng@vuw.ac.nz} }
\maketitle              
\begin{abstract}
Proteomics is the large-scale analysis of the proteins. The common method for identifying proteins and characterising their amino acid sequences is to digest the proteins into peptides, analyse the peptides using mass spectrometry and assign the resulting tandem mass spectra (MS/MS) to peptides using database search tools. However, database search algorithms are highly dependent on a reference protein database and they cannot identify peptides and proteins not included in the database. Therefore, \textit{de novo} sequencing algorithms are developed to overcome the problem by directly reconstructing the peptide sequence of an MS/MS spectrum without using any protein database. Current \textit{de novo} sequencing algorithms often fail to construct the completely matched sequences, and produce partial matches. In this study, we propose a genetic algorithm based method, GA-Novo, to solve the complex optimisation task of \textit{de novo} peptide sequencing, aiming at constructing full length sequences. Given an MS/MS spectrum, GA-Novo optimises the amino acid sequences to best fit the input spectrum. On the \mbox{testing} dataset, GA-Novo outperforms PEAKS, the most commonly used software for this task, by constructing 8\% higher number of fully matched peptide sequences, and 4\% higher recall at partially matched sequences.





\keywords{Genetic Algorithm, Tandem Mass Spectrometry, \textit{De Novo} Sequencing, Proteomics.}
\vspace{-3mm}
\end{abstract}

\section{Introduction}
\vspace{-2mm}

In mass spectrometry, \textit{de novo} peptide sequencing is the process of determining the amino acid sequence of peptides directly from MS/MS spectra. There are 20 amino acids represented by the letters A, C, D, E, F, G, H, I, K, L, M, N, P, Q, R, S, T, V, W, and Y. Peptide sequences are generally considered to be short chains of amino acids (from 2 to 50 amino acids). 
A peptide \textit{P} with length \textit{l} contains a sequence of amino acids, $P = a_1, a_2, a_3 ... a_l$, where each amino acid has a mass. 
Therefore, the mass of the peptide, which is called parent mass, equals to the total mass of its amino acids plus mass of water and is calculated based on the following equation \cite{frank2005pepnovo}.
\vspace{-3mm}

\begin{equation}
PM(P) = \sum\limits_{i=1}^{l} mass(a_i) + mass(H_2O)
\vspace{-1mm}
\label{eq:parent_mass}
\end{equation}

An MS/MS spectrum $ S $ consists of a list of peaks each having a mass-to-charge ratio (m/z) value and an intensity value (peak height). The m/z values are results of ionizing the biological samples and their intensities indicate the abundance of ions. Assume the spectrum is represented by two vectors of m/z values and intensities $S=(M, I)$, where $ M=({m}_{1},{m}_{2},{m}_{3},...,{m}_{n}) $ and \mbox{$I=(I_1,I_2,I_3,...,I_n) $}.
The experimental parent mass or precursor mass is calculated based on Equation \ref{eq:prec_mass}. 
\vspace{-3mm}

\begin{equation}
\mbox{Prec.$ _ {mass}$} = \mbox{pepmass $ \times $ charge - charge  $ \times $ mass(Proton)}
\label{eq:prec_mass}
\vspace{-1mm}
\end{equation}
where pepmass is mass of the fragmented ion, charge is the precursor charge state and and mass of Proton equals to 1.00727647 atomic mass units (amu).

Collision-induced dissociation (CID) is known to be highly suitable technique for the identification of peptide sequences \cite{papayannopoulos1995interpretation}. In this technique, fragmentation happens at the peptide bonds, producing b-/y-ions. 
The fragment containing only the first amino acid from left side (N-terminus) of the peptide is termed $b_1$, while the one that contains the first two amino acids is called the $b_2$ ion, and so forth. Y-ions extend from the right side or C-terminus of the peptide. In the CID fragmentation technique the amino acid sequence of an MS/MS spectrum can be determined by the mass differences between b- and y-ions.   

\begin{table} [t]
	\centering
	\caption{An example of a mass fragmentation ladder.}\label{table:ladder}
	\vspace{-3mm}
	\begin{scriptsize}
		\setlength{\tabcolsep}{10 pt}
		\begin{tabular}{|c l l l l l|} 
			\hline
			Mass & ion & b-ions & y-ions & ion & Mass \\ [0.5ex] 
			\hline
			114    & b1 & L &GVTLYK & y6 & 680 \\
			171  & b2 & LG & VTLYK & y5 & 623 \\
			270  & b3 & LGV & TLYK & y4 & 524 \\
			371  & b4 & LGVT & LYK& y3 & 423 \\
			484  & b5 & LGVTL & YK & y2 & 310 \\
			647  & b6 & LGVTLY & K& y1 & 147 \\[1ex] 
			\hline
		\end{tabular} 
	\end{scriptsize}
\vspace{-5mm}
\end{table}

The complete CID peptide fragmentation gives a contiguous series of ion types called \textquotedblleft ladder\textquotedblright \cite{wells2005collision}. Table \ref{table:ladder} shows an example of a mass fragmentation ladder for the peptide \textquotedblleft LGVTLYK\textquotedblright. It can be seen that each b-ion has a corresponding y-ion. These ions are called complementary ions when 
the sum of their masses equal to the mass of the pre-fragmented peptide. Having the complete ion ladder, the \textit{de novo} sequencing algorithm selects pairs of peaks and labels them if their mass differences are within the tolerance ranges of the \mbox{amino acid’s masses.}

However, it is often that peptide fragmentations are neither sequential nor complete. 
Moreover, peptides may not fragment at some positions and resulting in missing data. 
Also, a real MS/MS spectra with hundreds of peaks normally contain background noise.
Therefore, while exactly 1 of $ 20^\textit{l}$ amino acid sequences can be considered as the potential correct prediction (\textit{l} is the peptide length), \textit{de novo} sequencing with internal fragment ions is recognized as a combinatorial problem and known to be NP-hard \cite{xu2006complexity}. 

There have been attempts to solve the \textit{de novo} sequencing problem using different approaches. 
PAA3 \cite{sakurai1984paas}, as the first \textit{de novo} sequencing algorithm, generated exhaustively all possible peptide sequences and compared each candidate with the spectrum. However, the method is only feasible for very short peptides, because the time complexity grows exponentially in terms of peptide length. 
	
Dynamic programming has been widely used for \textit{de novo} sequencing \cite{ma2015novor}. The major approach is generating a graph from an MS/MS spectrum where peaks are the vertices and edges are defined as the corresponding amino acids to the mass differences between two vertices. 
A probability based fitness function is used to score the paths and dynamic programming is used to traverse through the spectrum graph \cite{ma2003peaks,bafna2003novo,dancik1999novo,nielsen2006characterization}. However, this approach results in having a huge graph due to the noise peaks caused by internal cleavages or post-translational modifications (PTM)s. Another problem is the lack of full path due to the missing ion types caused by incomplete fragmentation and low instrument accuracy. Therefore, \textit{de novo} sequencing of full-length peptides remains a challenge.
	
\textit{De novo} sequencing can be formulated as an optimisation problem where the objective is to discover the most likely amino acid sequence that can be generated by the input spectrum \cite{webb2007current}. 
\textit{De novo} sequencing has been performed via stochastic optimisation using a genetic algorithm (GA) \cite{heredia2004sequence,kistowski2011optimization}, where a GA tries to optimise the amino acid sequence in respect to a scoring function. However, these methods often fail to discriminate the mismatches because the fitness functions could not capture various aspects of peak matching \cite{allmer2011algorithms}. Moreover, the basic genetic operators used in these works are not capable enough to guide GAs during the evolutionary process to construct the fully matched sequence.
\vspace{-4 mm}
\subsection{Goal}
\vspace{-1 mm}
The goal of this work is to develop an effective \textit{de novo} sequencing algorithm using GAs to construct the full length amino acid sequences of MS/MS spectra.  
Unlike exhaustive approaches, GA does not need to generate all possible amino acid sequences for a give spectrum. A set of initial amino acid sequences using an effective initialisation method is generated and during the evolutionary process these sequences are manipulated by appropriate domain dependant genetic operators until finding the one that best fits to the spectrum in respect to the fitness function. Unlike spectrum graph based algorithms, it is expected that the performance of GA does not deterred by discontinuities in the search space (lack of full path in the graph) due to missing ions. 
Therefore, the following \mbox{objectives} are investigated in this work:
\vspace{-2mm}
\begin{enumerate}
	\item Developing a new fitness function that captures important spectral features and enables GA to discriminate the mismatches.
	
	\item Developing an effective set of mutation and crossover operators that help GA to construct the full length amino acid sequence.

	\item Designing an effective GA algorithm that can perform the \textit{de novo} sequencing task, and achieving a high number of fully matched sequences out of the input spectra.   
\vspace{-2mm}
\end{enumerate}


\vspace{-4mm}
\section{The proposed Method}
\begin{figure}[t]
	\centering
	\includegraphics[width=\columnwidth]{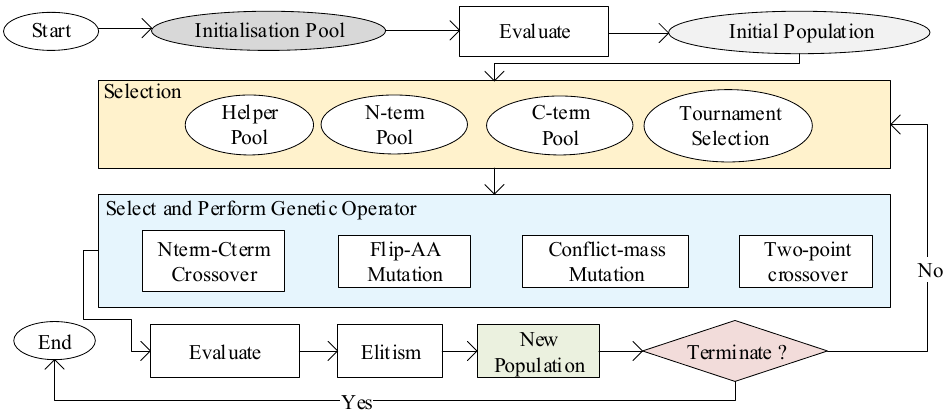}\vspace{-5mm}
	\caption{The workflow of GA-Novo.}
	\label{fig:GA_Novo_workflow}
	\vspace{-5mm}	
\end{figure}

\vspace{-2mm}
Fig. \ref{fig:GA_Novo_workflow} presents the workflow of GA-Novo. Given the raw MS/MS experimental spectrum \textit{S}, first a tag-based initialisation method is applied in order to create a set of candidate initial individuals for the GA algorithm. The candidate individuals are kept in a big initialisation pool. The individuals are evaluated and based on three criteria including the fitness value, Nterm score and Cterm score are selected to generate the initial population for GA. Then the evolutionary process starts with applying selection in order to create four pools for different purposes and the size of each pool is a third of the total population size.
Starting from left, the helper pool contains top best individuals in terms of fitness values. The individuals in N-term and C-term pools are the top best individuals in terms of Nterm and Cterm scores, respectively. The individuals in last pool are selected using tournament selection based on their fitness values. There are four genetic operators, two crossovers and two mutations. The individuals for Nterm-Cterm crossover are selected from the first three pools. Other genetic operators get their individuals directly from the tournament pool. Nterm-Cterm crossover is designed to construct individuals with correct matches from both sides and possibly from middle, whereas two-point crossover aims to repair the individuals from middle. The mutation operators randomly flip flop the each bit/amino acid in the sequence. In each generation, elitism keeps the best three individuals in terms of overall fitness value, Nterm and Cterm score. The evolutionary process repeats until the termination criterion which is the number of generations is met. The method returns the best individual in terms of overall fitness value. More details about the components in this flowchart are as follow.

\vspace{-3mm}
\subsection{Representation}
\vspace{-1mm}
Each GA individual is variable-length and represented by a sequence of single-letter amino acids from, for example \textquotedblleft AAALAAADAR\textquotedblright. 
Each individual contains three fitness scores including the overall fitness value (from the fitness function in Equation \ref{eq:fitness_function}), Nterm and Cterm scores which are explained later. 
\vspace{-4mm}
\subsection{Tag-based Initialisation Method}
\begin{figure}[t]
	\centering
	\includegraphics[width=\columnwidth]{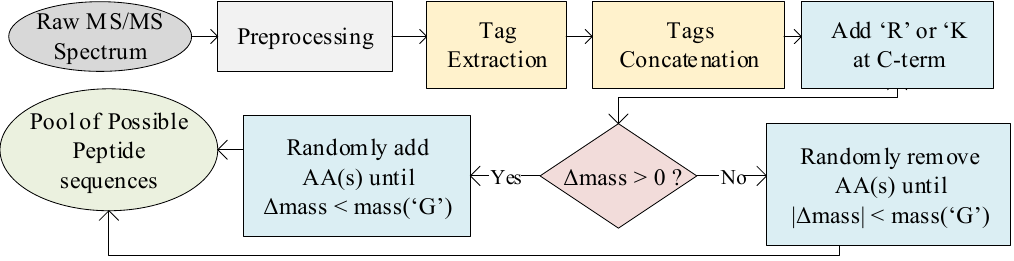}\vspace{-4mm}
	\caption{The workflow of the tag-based initialisation method.}
	\label{fig:initialisation}
	\vspace{-5 mm}
\end{figure}

\vspace{-1mm}
A domain dependant initialisation method is used to generate initial individuals for GA. 
The workflow of this method is illustrated in Fig. \ref{fig:initialisation}. The overall goal of this method is to construct full length peptide sequences which are preferably partially matched with the spectrum and having as small as possible mass difference ($ \Delta mass $). 

The input of the workflow is an MS/MS spectrum (experimental spectrum) and the output is a set of peptide sequences corresponding to the spectrum. The workflow starts with preprocessing the input spectrum. Then all 3-letter tags are extracted from the preprocessed spectrum in tag extraction step. In tags concatenation step, each time 2, 3 or 4 tags are randomly selected and concatenated to construct a sequence with length 6, 9 or 12. These numbers are in the range of the peptides' length that fall in the precursor mass range of spectra used in this study. Since all tryptic peptides have either amino acids \textquoteleft R' or \textquoteleft K' at the end, these two amino acids are randomly added to the end of the sequences from tags concatenation step. Since mass difference is a constraint, it is important to construct the sequences with $ |\Delta masses| \leq 0$. So the rest of the workflow checks whether or not the mass difference between each constructed sequence and the spectrum is less than the mass of amino acid \textquoteleft G' ~which has the smallest mass. 
Therefore, based on the $ \Delta mass $ value, appropriate amino acids are randomly added to/removed from the sequence and the resulting peptide sequence is sent to the pool of possible peptide sequences corresponding to the input experimental spectrum. The preprocessing step and the tag extraction are explained in the following.
\vspace{-5mm}
\subsubsection{Spectrum Preprocessing.}
The MS/MS noise reduction step has been done based on the noise reduction method proposed in SEQUEST \cite{eng1994approach}, which is a dominant database search tool in proteomics. 
Given a spectrum, at first the whole m/z range is divided into 10 windows (regions). In each window, if the number of existing peaks exceeds 9, there should be some possible noise, which needs to be eliminated from that window. 
The peak intensity with the highest frequency is considered to be the noise threshold. Therefore, all peaks whose intensities are smaller than the noise threshold will be removed from that window.
After removing these noisy peaks, the next step is normalising peak intensities. In each window, each peak\textquoteright s intensity is replaced with its square root and then all intensities are normalised by dividing into the highest intensity. 
Then each peak in the spectrum is checked for the existence of its complementary peak which will be added if required. 
The sum of the two complementary ions\textquoteright~masses should be equal to the precursor mass of the spectrum. 
Now the next step is extracting all 3-letter tags from the spectrum. 
\vspace{-5mm}
\subsubsection{Tag extraction.}
In tag extraction, all 3-letter tags from the N-terminus to the C-terminus are extracted from the spectrum \cite{yu2016pipi}. As previously mentioned, here a spectrum is represented by two vectors of mz values and intensities $S = (M, I)$. Considering the M vector $ M=({m}_{1},{m}_{2},{m}_{3},...,{m}_{n}) $, two peaks construct a peak pair if their $m/z$ values satisfy $|{m}_i-{m}_j-mass(a)|\leq\tau$ where $ 1\leq i\leq j\leq n $, $mass(a)$ is the mass of one of the 20 popular amino acids and $\tau$ is the MS/MS mass tolerance. A tag with length one is represented by $t(i,j)$ and a label of $a$ corresponding to its amino acid. Two tags $t(i,j)$ and $t(i\prime,j\prime)$ are considered sequential if $ j = i\prime$. So all 3-letter tags from the spectrum will be extracted and are used in the initialisation method.
\vspace{-4mm}
\subsection{Fitness}
\vspace{-1mm}
\subsubsection{Fitness Function.}
The fitness function evaluates the quality of matching between an input experimental spectrum and a peptide sequence constructed by GA-Novo. For being able to match the peptide sequence against the experimental spectrum, a theoretical spectrum \textit{T} based on the known CID fragmentation rules of doubly charged peptides \cite{herrmann2006peptide} is constructed from the peptide sequence. The theoretical spectrum only contains m/z values with no intensities. Both b-/y-ion ladders in Table \ref{table:ladder} along with internal fragments are constructed in the theoretical spectrum. 
Then each peak in the theoretical spectrum is matched against the peaks in the experimental spectrum within the MS/MS mass tolerance of $\tau$.

Equation \ref{eq:fitness_function} presents the new fitness function for measuring the goodness of the peptide spectrum match (PSM). 
\begin{equation}
\small
fitness(PSM) = \frac{{\sum\limits_{}^{} I_{matched}}}{\sum\limits_{i=1}^{n} I_i} - \frac{|\Delta mass|}{\mbox{Prec.$ _ {mass}$}} + \frac{ Nterm + Cterm - \sum\limits_{}^{} {N_{unmatched}}}{length(P)} 
\label{eq:fitness_function}
\vspace{-2mm}
\end{equation}
where $ I_{matched} $ is the sum of intensities of those peaks in the experimental spectrum \textit{S} which are matched with theoretical spectrum \textit{T} corresponding to the peptide \textit{P}. 
Then total intensities of matched peaks is normalised by dividing into the total intensities of the whole spectrum \textit{S}. $\Delta mass$ is the mass difference between parent mass of peptide \textit{P} and the spectrum precursor mass ({\mbox{Prec.$ _ {mass}$}}). Since the total mass of the predicted peptide by GA is expected to be equal to the precursor mass of the spectrum, the absolute value of $\Delta$mass is considered as a penalty to avoid getting undesirable short or long peptides. $ Nterm $ is the number of sequential b-ion matches from N-terminus (left to right) and $ Cterm $ is the number of sequential y-ion matches from C-terminus (right to left) of the theoretical spectrum \textit{T}. These terms check the quality of match from both sides of the theoretical spectrum and reward the match. As normally those b-/y-ions in the middle part of the spectrum tend to have higher intensities, whereas those on the other two sides particularly N-terminus have lower intensities, without having these two terms in the fitness function there is a chance of ending up to a peptide sequence which is partially matched with the spectrum only from middle. Therefore, with having these two terms, a peptide which has a few b-/y-ions matched from two sides but not from middle, still has the chance to survive. In this case, the peptide gets a reasonable fitness value and has a chance to remain in the population, going through the evolutionary process for further improvement. $N_{unmatched}$ indicates the number of b-/y-ions in the theoretical spectrum \textit{T} which are not a match against the spectrum \textit{S}. The three terms are divided into the length of peptide.

Apart from the fitness value produced by the fitness function above, the two terms $ Nterm $ and $ Cterm $ (without being divided into the peptide length), are also kept as additional fitness scores for each individual. These values later are used to apply a new crossover operator and are explained in the following section.
\vspace{-5mm}
\subsubsection{ Nterm and Cterm scores.}  
The idea of calculating these two terms comes from the ion ladder of sequences and the CID fragmentation rules. The mass of any theoretical b-ion can be calculated based on Equation \ref{eq:b-ion},
where $ 1\leq j\leq l-1 $, \textit{l} is the length of the peptide \textit{P}, $ b_j $ is the j\textit{-th} b-ion of $ \textit{P} $, and $a_i$ is the i\textit{-th} amino acid in \textit{P}. Similarly theoretical y-ions can be calculated based on Equation \ref{eq:y-ions}. Also as mentioned in Table \ref{table:ladder}, the complementary theoretical b- and y- ions in each row of the table have the mathematical relation presented in Equation \ref{eq:b-y-ions}.
\vspace{-3mm}
\begin{equation}
\vspace{-2mm}
b_j = {\sum\limits_{i=1}^{j} mass(a_i)} + 1
\label{eq:b-ion}
\vspace{-2mm}
\end{equation} 
\begin{equation}
y_j = {\sum\limits_{i=l-j}^{l} mass(a_i)} + 19
\label{eq:y-ions}
\vspace{-2mm}
\end{equation} 
\begin{equation}
\vspace{-1mm}
b_j + y_{l-j} = PM(P) + 2
\label{eq:b-y-ions}
\end{equation} 

To calculate the b-ions in the theoretical spectrum Equation \ref{eq:b-ion} is used. Having the total mass of the peptide (parent mass in Equation \ref{eq:parent_mass}), the y-ions can be calculated either by Equation \ref{eq:b-y-ions} or Equation \ref{eq:y-ions}. Therefore, for calculating the Nterm score, fist all b-ions are calculated. Then, in Equation \ref{eq:b-y-ions} instead of $ PM(P)$ which the mass of the peptide, \mbox{Prec.$ _ {mass}$} which is the precursor mass (from Equation \ref{eq:prec_mass}) is replaced, and y-ions are calculated. Let\textquoteright s call these y-ions as experimental y-ions (becasue we used mass of the spectrum). The experimental y-ions are compared with theoretical y-ions from \mbox{Equation \ref{eq:y-ions}}. Starting from $ y_1 $, if any two sequential experimental y-ions are equal to the corresponding theoretical y-ions, the Nterm score increases by one. 

The Nterm score is able to check whether a matched b-ion is a random match or not. Similarly, Cterm is calculated by using Equation \ref{eq:y-ions} and applying the similar process. The values of Nterm and Cterm scores do not necessarily indicate the exact amino acid matches in the sequence. For example a sequence with Nterm = 2, does not indicate that the first 2 amino acids from N-terminus are exact matches compared to the ground truth peptide. The reason is that we are not aware of the ground truth during the matching process. However, these two scores are able to check the quality of match from each side of the spectrum, and check whether it is a random match from one side or a potential correct match from two sides of the spectrum.



\vspace{-4mm}
\subsection{Nterm\mbox{-}Cterm Crossover}
\vspace{-1mm}
\begin{figure}[t]
	\centering
	\includegraphics[width=\columnwidth]{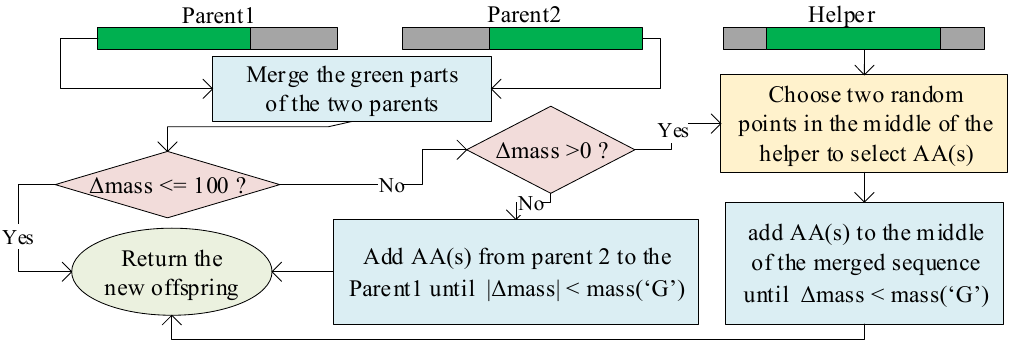}\vspace{-5mm}
	\caption{The workflow of Nterm\mbox{-}Cterm crossover operator.}
	\label{fig:crossover}
\end{figure}

A new domain specific crossover is designed for this problem. The crossover mates two parents each having at least one exact match one b-/y-ion from N-terminus and C-terminus. The goal is to mate these two parents in the way that the new offspring would have exact b-/y-ion matches from both sides and possibly from the middle as well. Fig. \ref{fig:crossover} illustrates the Nterm\mbox{-}Cterm crossover workflow. The input of the crossover is three GA individuals, two individuals as parents and one as a helper, and the output is a new offspring. At first, the exact match parts (the green parts) from both parents are concatenated. Here the $\Delta mass$ condition is more relaxed, allowing up to 100Da mass difference. If the $\Delta mass$ is more than absolute value of 100, then the new concatenated sequence is checked whether it needs to remove/add amino acids from/to the sequence. A negative $\Delta mass$ indicates that the sequence is long and needs removing a few amino acids from it and vice versa for a positive value. The reason is that based on Equation \ref{eq:parent_mass} a long sequence has more amino acids and possibly it could have a bigger parent mass compared to a shorter sequence with less number of amino acids. Since there might be some overlap between the green parts of the two parents, these $ \Delta mass$ conditions help the operator to avoid constructing a bad offspring having a big $\Delta mass$ penalty in its fitness value. Therefore, when $\Delta mass$ is negative, for removing the overlap the green part of the parent 1 is considered as the N-terminus of the new sequence and each time one amino acid from C-terminus (the most right) of the parent 2 is added to the new sequence until the $ \Delta mass $ criterion is met. 

If $ \Delta mass $ is positive, it is required to add a few amino acids in the middle of the green parts of the two parents. Here instead of adding random amino acids, another individual as helper is used. The helper parent has a high fitness value which possibly could indicate having more matched peaks in the middle. So two crossover points are picked randomly from the middle of helper parent and the amino acids in between those two points are added to the middle of the new sequence one by one until the mass difference criterion is met. 





\vspace{-4mm}
\subsection{Flip-AA Mutation}
\vspace{-1mm}
The flip-AA mutation randomly pick one amino acid from the sequence and replaced it with one of the 19 amino acids (\textquoteleft I' and \textquoteleft L' are considered identical). The mutation operator is not allowed to mutate the last amino acid in the sequence as it is always supposed to be either \textquoteleft R' or \textquoteleft K' ~for a tryptic peptide sequence.
\begin{table} [t]
	\centering
	\caption{The dictionary of conflict masses.} 
	\vspace{-3mm}
	\label{table:conflict_masses}
		\setlength{\tabcolsep}{8pt}
		\begin{tabular}{|c l l |} 
			\hline
			single AA & di-peptide  & mass \\ [0.5ex] 
			\hline
			W & DA, AD, EG, GE, VS, SV 	 & 186  \\
			R & VG, GV & 156 \\
			Q & AG, GA  & 128  \\
			N & GG  & 114 \\ 
			\hline
		\end{tabular} 
\end{table}
\vspace{-4mm}
\subsection{Conflict-mass Mutation}
\vspace{-1mm}

There are situations where the mass of a single amino acid conflicts with the mass of two amino acids (di-peptides). For example, the mass(\textquoteleft W') = mass (\textquotedblleft DA") \mbox{= 186}. A dictionary of such conflict masses is provided and shown in Table \ref{table:conflict_masses}. The conflict-mass mutation operator checks whether the sequence contains any amino acid in the conflict mass dictionary and randomly replaces the amino acid with any of the corresponding di-peptides.

\vspace{-4mm}
\section{Experiment Design}
\vspace{-1mm}
\subsection{Dataset}
\vspace{-1mm}
The comprehensive full factorial LC-MS/MS benchmark dataset, which is particularly designed for evaluating MS/MS analysis tools, containing 50 protein samples extracted from Escherichia coli K12, is used in this study \cite{wessels2012comprehensive}. The dataset was acquired from the linear ion trap Fourier-transform (LTQ-FT, Thermo Fisher Scientific) with the collision-induced dissociation (CID) technique. The MS/MS spectra in this dataset have been already searched against a curated Refseq \cite{maglott2000ncbi} release 33 \textit{E.coli} database by using Mascot v2.2 \cite{cottrell1999probability}. 
From the peptide identification results provided by this dataset, a set of 120 doubly charged peptide-spectrum matches (PSMs) with a minimum Mascot peptide identification score of 45, minimum peptide length of 7 amino acids and maximum length of 12 is selected. Based on Table \ref{table:dataset}, the average length of the peptides is 9.5. The spectra have a precursor of less than 1150 Da and the fragment ion of 0.5 is used as the value of tolerance $\tau$. The so-called \textquotedblleft ground truth" is used to test the performance of \textit{de novo} sequencing algorithms.

\begin{table} [t]
\vspace{-1mm}	
	\centering
	\caption{The set of peptide spectrum matches used in this study.} 
	\vspace{-3mm}\label{table:dataset}
		\setlength{\tabcolsep}{3pt}
			\begin{tabular}{|c|c|c|c|c|c|}
				\hline
				\begin{tabular}[c]{@{}c@{}}no. of \\ PSMs\end{tabular} & \begin{tabular}[c]{@{}c@{}} peptide\\ length range\end{tabular} & \begin{tabular}[c]{@{}c@{}}avg. length\\ of peptides\end{tabular} & Charge No. & \begin{tabular}[c]{@{}c@{}}Precursor\\ mass range\end{tabular}& fragment ion (Da) \\ \hline
				120                                                          & 7-12                                                             & 9.5                                                                  & 2          & \textless{}1150     &  0.5                                           \\ \hline
			\end{tabular} 
\end{table}

	\begin{table}[t]
	\vspace{-1mm}
	\caption{GA-Novo parameters}

	\begin{center}
			\vspace{-3mm}
		\begin{tabular}{|l|c||l|c|}
			\hline
			Parameter & Value & Parameter & Value \\ \hline
			Initialisation Pool Size & 1000 & Population Size & 300 \\ \hline
			Size of Sub-pools & 100 & Generations, Runs & 50, 30 \\ \hline
			Flip-AA Mutation Rate & 0.1 & Conflict-mass Mutation Rate & 0.15 \\ \hline
			Crossover Rate (2point) & 0.35 & Elitism Rate & 0.01  \\ \hline
			Nterm-Cterm Crossover Rate & 0.40 & Selection & Tournament, 7 \\ \hline
		
		\end{tabular}
		\label{table:GA_parameters}
	\end{center}
	\vspace{-9mm} 
\end{table}

\vspace{-4mm}
\subsection{Parameters, Evaluation and Benchmark Algorithm}
\vspace{-1mm}
The parameters in Table \ref{table:GA_parameters} are used to setup the GA algorithm. A-Novo is implemented in Python 3.6 and uses DEAP (Distributed Evolutionary Algorithms in Python) package \cite{DEAP_JMLR2012}. To evaluate the accuracy of \textit{de novo} sequencing results, the \textit{de novo} peptide sequences constructed by the algorithm are compared with the real peptide sequences from the ground truth dataset. The total recall and precision metrics are calculated based on the following equations:

\begin{equation}
\mbox{precision} = {\frac{ \mbox{total number of matched amino acids}}{ \mbox{total length of predicted peptide sequences}}}
\label{eq:evaluation_precision}
\end{equation}

\begin{equation}
\mbox{recall} = {\frac{ \mbox{total number of matched amino acids}}{ \mbox{total length of ground truth peptide sequences}}}
\label{eq:evaluation_Recall}
\end{equation}
The performance of GA-Novo is compared with PEAKS, which is a popular benchmark \textit{de novo} sequencing algorithm \cite{ma2003peaks}. The metrics in both \mbox{Equation \ref{eq:evaluation_precision}} and Equation \ref{eq:evaluation_Recall} measure the accuracy of the results in amino acid level. The following metric is also used to evaluate the results of both algorithms in peptide level.
\begin{equation}
\mbox{recall}_{peptide~level} = {\frac{ \mbox{total number of fully correctly predicted peptide sequences}}{ \mbox{total number of ground truth peptides}}}
\label{eq:evaluation_peptide_level}
\vspace{-7mm}
\end{equation}

\section{Results and Discussions}
This section presents three different experiments. The first experiment uses GA-Novo  for \textit{de novo} sequencing of 120 MS/MS spectra in the dataset and the results are compared with those of PEAKS. The rest of this section analyses the effectiveness of two main components used in GA-Novo namely tag-based initialisation method and the domain dependant Nterm-Cterm crossover. Therefore, the second experiment compares random and tag-based initialisation methods followed by the third experiment that evaluates the effectiveness of Nterm-Cterm crossover and gives two examples of how this operator can help GA-Novo to construct the fully matched sequences.
\vspace{-4mm}
\subsection{Performance Comparison Between GA-Novo and PEAKS} 
This section compares the overall performance of GA-Novo with PEAKS. All spectra in the dataset (Table \ref{table:dataset}) are used to assess the performance of both algorithms. Among these spectra some of them are noisy and some might have incomplete ion ladders. 

Given an MS/MS spectrum to PEAKS, the output is a set of peptide sequences each having a confidence score level between 0 and 100 \cite{ma2003peaks}. The score indicates how likely the complete sequence is correct. For each spectrum, the top scored sequence is taken as the output of \textit{de novo} sequencing by PEAKS. PEAKS was run with an error tolerance of 0.5 Da and tryptic digestion. 

For GA-Novo, the experiments are repeated for 30 independent runs with 30 different random seeds. For each spectrum in each run, the best fit sequence constructed by the GA algorithm is taken as the output of GA-Novo. To compare the results of GA in 30 runs with PEAKS, one sample statistical t-test with 95\% confidence interval is used to compare the performance of two methods.   
Table \ref{table:GA-Novo_120_Spectra} presents the results of \textit{de novo} sequencing by these two methods. (+) in the table indicates the difference between the results of GA-Novo and PEAKS is considered to be statistically significant and (=) indicates not statistically significant. 

\vspace{-1mm}

\begin{table} [t]
	\vspace{-1mm}
	\centering
	\caption{The results of sequencing 120 MS/MS spectra by GA-Novo and PEAKS.}
	\vspace{-3mm}\label{table:GA-Novo_120_Spectra}
	\setlength{\tabcolsep}{3 pt}
\begin{tabular}{|c|c|c|c|c|c|}
	\hline
	Algorithm & Precision & Recall & $ \mbox{recall}_{pep.~level} $ & \begin{tabular}[c]{@{}c@{}}avg. len. of \\ partial matches\end{tabular} & \begin{tabular}[c]{@{}c@{}}avg. len. of  \\ predicted\\ sequences\end{tabular} \\ \hline
	GA-Novo & \begin{tabular}[c]{@{}c@{}}0.89 $ \pm $ 0.03\\ \textbf{($ \textbf{+}$)}\end{tabular} & \begin{tabular}[c]{@{}c@{}}0.88 $ \pm $ 0.03\\ \textbf{($ \textbf{+}$)}\end{tabular} & \begin{tabular}[c]{@{}c@{}}0.64 $ \pm $ 0.06\\ \textbf{($ \textbf{+}$)}\end{tabular} & \begin{tabular}[c]{@{}c@{}}8.4 $ \pm $ 0.27\\ \textbf{($ \textbf{+}$)}\end{tabular} & \begin{tabular}[c]{@{}c@{}}9.4 $ \pm $ 0.1\\ (=)\end{tabular} \\ \hline
	PEAKS & 0.85 & 0.84 & 0.56 & 8.06 & 9.43 \\ \hline
\end{tabular}
	\vspace{-5mm}
\end{table}

%
%
It can be seen that the results of GA-Novo in most cases are statistically significant. GA-Novo outperforms PEAKS by 4\% increase in precision and 4\% increase in recall. Moreover, the accuracy of fully matched peptide sequences predicted by GA-Novo, recall in peptide level, is 8\% higher than PEAKS. The reason of having lower recall compared to the precision in the results of both algorithms is that, they mainly construct either equal or slightly shorter peptide compared to the real peptide in terms of length. Also, the results show that in overall GA-Novo is able to find more partially matched sequences compared to PEAKS, as the average length of partially matched sequences for GA-Novo is 8.4 and statistically significant than the result of PEAKS.

As shown in Table \ref{table:dataset} that the average length of peptides in this dataset is 9.5, sequences predicted by GA-Novo and PEAKS have the average length of about 9.4. No doubt that this value is close to the average length of the peptides in ground truth as the goal of both algorithms is constructing full length individuals.

The sequences \textquotedblleft AMVEVFLER" and \textquotedblleft DAGTLLWLGK" are two examples of when PEAKS failed to predict the whole sequences , whereas GA-Novo could successfully construct the fully matched peptides. The sequences were predicted by PEAKS as \textquotedblleft TT\textbf{VEVFLER}" and \textquotedblleft W\textbf{GTLLWLGK}" while the first two amino acids in both sequences were predicted wrongly. More analysis on the results of PEAKS shows that it sometimes fails to predict the conflict masses from Table \ref{table:conflict_masses}, whereas GA-Novo gets benefit of its domain dependant mutation operator, conflict-mass mutation to avoid these types of mismatches.

Although the results show that GA-Novo is able to construct the full length of sequences (9.4 relatively close to 9.5), GA-Novo also sometimes fails to construct the fully matched sequences (8.4). However, comparing the difference between the average length of ground truth peptides, 9.5, and the results of average length of partial matches for GA-Novo, 8.4, the result shows that in overall GA only fails to fully match either one or two amino acids. The reason of this mismatch is the conflict mass of di-peptides. As mentioned previously in Table \ref{table:conflict_masses} where the mass of di-peptides conflicts with the mass of one single amino acid, there are other situations where the mass of two di-peptides conflict with each other.

\vspace{-2mm}
\subsection{Tag-based Initialisation vs. Random initialisation}
\vspace{-1mm}
\begin{figure}[t]
	\centering
	\includegraphics [width=\columnwidth]
	{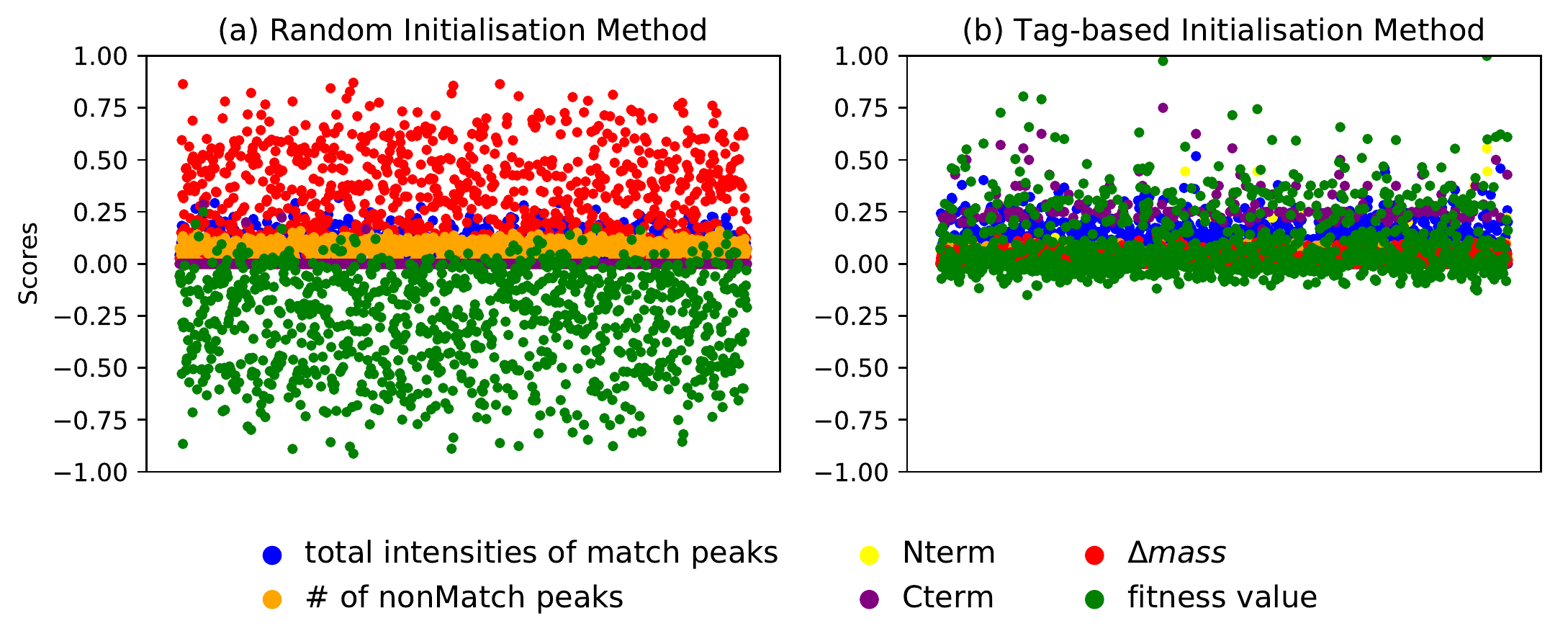}\vspace{-5mm}
	\caption{Plots of 1,000 individuals generated by random and tag-based initialisation.}
	\label{fig:initial_vs_random}
	\vspace{-3mm}
\end{figure}
Fig. \ref{fig:initial_vs_random} illustrates two plots presenting the overall fitness value and the values of its 5 terms included in the fitness function (see Equation \ref{eq:fitness_function}). As the random initialisation method does not use any domain knowledge and randomly generates sequences between length 7 and 12, it can be seen from Fig. \ref{fig:initial_vs_random}(a) that the fitness values of majority of population is below zero. The reason of such low fitnesses is that the random initialisation does not pay attention to $\Delta mass$, mass difference, which is a penalty in fitness function. Generating short or long peptide sequences results in a big $\Delta mass$ penalty. However, the tag-based initialisation plot in this figure, shows how the $ \Delta mass $ values are small in this method and the overall fitness values are bigger than random method. 

The results in Table \ref{table:stat-initial_single} show that the best individual out of 1000 individuals in a single run of random method is \textquotedblleft YVMNEAR" with a fitness value of 0.25. In this table, each sequence is shown by its overall fitness value and fine different terms from fitness function, including I, D and N which indicate the total intensities of matched peaks, $ \Delta mass $ and the number of unmatched peaks, respectively. Nterm and Cterm are normalised here. Based on the tag-based method, the best individual is \textquotedblleft RVAAAAWR" with fitness value of 1.14. Therefore, the fitness value of the best individual produced by tag-based initialisation method is 4.7 times bigger than the one in random initialisation. As mentioned above the fitness value of the ground truth is 2.19, therefore the tag-based initialisation method could be a better start point for GA.

The statistics results in Table \ref{table:stat-initial_30runs} show the significance of comparison between the results of two methods. An unpaired statistical t-test with 95\% confidence interval is used to compare the performance of two methods. In this table, (+) indicates a significant change and (=) indicates no difference. This table presents the statistics on overall fitness value, Nterm and Cterm scores. Please notice that Nterm and Cterm scores are not normalised here. From this table it can be seen that the average fitness values and Cterm scores of tag-based initialisation method are statistically significant than random based method. However, Nterm scores are not statistically significant than random method, thanks to the tag extraction step which sometimes is not able to extract partially matched 3-letter tags from N-terminus of the spectrum due to the missing b-ions in this area. During peptide fragmentation, peptides may not fragment at some positions and leave no information, resulting in missing data. That is why the first two fragments $ b_1 $ and $ b_2 $ ions are seldom observed in the spectrum.


\begin{table} [t]
	\vspace{-1mm}
	\centering
	\caption{The best individual in a single run tag-based and random initialisation methods using the spectrum of \textquotedblleft AAALAAADAR\textquotedblright peptide.}
	\vspace{-3mm}\label{table:stat-initial_single}
	\setlength{\tabcolsep}{1 pt}
	
	\begin{tabular}{cc|c|c|c|c|c|c|}
		\cline{3-8}
		&  & \multicolumn{6}{c|}{Fitness scores} \\ \cline{2-8} 
		\multicolumn{1}{c|}{} & Sequence & ~Fitness & ~I & ~D & ~N & ~Nterm & ~Cterm \\ \hline
		\multicolumn{1}{|c|}{Ground-Truth} & AAALAAADAR & 2.1950 & 0.595 & 0.000003 & 0.0 & 0.8 & 0.8 \\ \hline \hline
		\multicolumn{1}{|c|}{\begin{tabular}[c]{@{}c@{}}Random Initial.\end{tabular}} & YVMNE\textbf{AR} & 0.25 & 0.057 & 0.020099 & 0.071 & 0 & 0.28 \\ \hline
		\multicolumn{1}{|c|}{\begin{tabular}[c]{@{}c@{}}Tag-based Initial.\end{tabular}} & RVA\textbf{AAA}W\textbf{R} & \textbf{1.14} & 0.528 & 0.000027 & 0.0 & 0 & 0.625 \\ \hline
	\end{tabular}
\end{table} 
\begin{table} [t]
	\vspace{-1mm}
	\centering
	\caption{The statistics on three fitness scores in 30 different runs of tag-based and random initialisation methods using the spectrum of \textquotedblleft AAALAAADAR\textquotedblright peptide.}
	\vspace{-3mm}\label{table:stat-initial_30runs}
	\setlength{\tabcolsep}{2 pt}
	
	\begin{tabular}{l|c|c|c|c|c|c|c|c|c|c|c|c|}
		\cline{2-13}
		& \multicolumn{4}{c|}{Fitness value} & \multicolumn{4}{c|}{Nterm} & \multicolumn{4}{c|}{Cterm} \\ \cline{2-13} 
		& Min & Max & Avg. & Std. & Min & Max & Avg. & Std. & Min & Max & Avg. & Std. \\ \hline
		\multicolumn{1}{|l|}{\begin{tabular}[c]{@{}l@{}}Random Initial.\end{tabular}} & -0.97 & 0.32 & -0.29 & 0.24 & 0 & 1.57 & 0.002 & 0.06 & 0 & 2.57 & 0.01 & 0.15 \\ \hline
		\multicolumn{1}{|l|}{\begin{tabular}[c]{@{}l@{}}Tag-based Initial.\end{tabular}} & -0.15 & 1.07 & 0.1 & 0.17 & 0 & 4.83 & 0.04 & 0.35 & 0 & 5.93 & 0.45 & 0.98 \\ \hline
		\multicolumn{1}{|c|}{Significance} &
		 \multicolumn{4}{c|}{\textbf{($ \textbf{+}$)}}&  
		 \multicolumn{4}{c|}{(=)} &
		 \multicolumn{4}{c|}{\textbf{($ \textbf{+}$)}}  \\ \hline
	\end{tabular}
	\vspace{-5mm}
\end{table}
In overall based on the results in both Tables \ref{table:stat-initial_single} and Table \ref{table:stat-initial_30runs}, tag-based method constructs better/fitter individuals compared to random initialisation, as tag-based method focuses on concatenating randomly 2, 3 or 4 tags. Then the method reduces the absolute mass differences between the constructed sequences and the spectrum by randomly inserting/removing random amino acids into the sequences. As a known domain knowledge, each tryptic peptide ends in either \textquoteleft K' or \textquoteleft R', so this heuristic has been applied randomly on the sequences constructed by this method as well. As the result this method decreases the mass differences and increases the number of match ions, resulting in an increase in the total intensities of the match ions. Back to the best sequence produced by the tag-based initialisation, \textquotedblleft RVAAAAWR" in Table \ref{table:stat-initial_single}, it is expected this sequence goes through the GA evolutionary process and after a few generations converts to the exact match.

\vspace{-4mm}
\subsection{ Analysis the Effectiveness of Nterm-Cterm Crossover}
\vspace{-1mm}
This section presents two examples when Nterm-Cterm Crossover is applied on different individuals and also shows the performance of this operator across 30 different runs. Table \ref{table:CX_long_parents} and Table \ref{table:CX_short_parents} show how new Nterm-Cterm Crossover can result in whole sequence exact match. By looking at the Nterm and Cterm scores of Nterm and Cterm parents in Table \ref{table:CX_long_parents}, it can be seen that these parents have quite big values that could indicate potential exact amino acid matches from each side. Considering the sequence of amino acids of these parents and knowing the ground truth, it can be seen that the two parents have a few number of exact amino acid matches. However, concatenating the exact match amino acids (shown in bold), results in a false sequence \textquotedblleft AAALALAAADAR" which is not desired. As the technique was explained previously, the two parents are concatenating with consideration of removing the overlap and this results in a whole sequence exact match as the offspring. As both parents have enough Nterm and Cterm match amino acids, the third parent, helper, is not used here.
\begin{table} [t]
	\vspace{-1mm}
	\centering
	\caption{An example of applying Nterm-Cterm crossover on two long partially matched parents that have matched amino acids overlap.}\label{table:CX_long_parents}
	\vspace{-3mm}
	\setlength{\tabcolsep}{3.5 pt}
	\begin{tabular}{l|l|c|c|c|c|c|c|}
		\cline{2-8}
		& Sequence & \multicolumn{1}{l|}{fitness} & \multicolumn{1}{l|}{I} & \multicolumn{1}{l|}{D} & \multicolumn{1}{l|}{N} & \multicolumn{1}{l|}{\begin{tabular}[c]{@{}l@{}}Nterm\\ score\end{tabular}} & \multicolumn{1}{l|}{\begin{tabular}[c]{@{}l@{}}Cterm\\ score\end{tabular}} \\ \hline
		\multicolumn{1}{|l|}{$Nterm_{parent}$} & \textbf{AAALA}GGWR & 0.79 & 0.21 & 0.031 & 0.05 & 4 & 2 \\ \hline
		\multicolumn{1}{|l|}{$Cterm_{parent}$} & NV\textbf{LAAADAR} & 1.34 & 0.58 & 0.000002 & 0.02 & 0 & 7 \\ \hline
		\multicolumn{1}{|l|}{$Helper_{parent}$} & RG\textbf{LAAAD}VK & 0.58 & 0.59 & 0.00003 & 0.01 & 0 & 0 \\ \hline
		\multicolumn{1}{|l|}{$Offspring$} & \textbf{AAALAAADAR} & 2.19 & 0.59 & 0.000003 & 0.000 & 8 & 8 \\ \hline
	\end{tabular}
	\vspace{-2mm}
\end{table}
\begin{table} [t]
	\vspace{-0.5mm}
	\centering
	\caption{An example of applying Nterm-Cterm Crossover on two short partial match parents and a helper parent to fill the middle gap.}\label{table:CX_short_parents}
	\vspace{-3mm}
	\setlength{\tabcolsep}{2 pt}
	\begin{tabular}{l|l|c|c|c|c|c|c|}
		\cline{2-8}
		& Sequence & \multicolumn{1}{l|}{fitness} & \multicolumn{1}{l|}{I} & \multicolumn{1}{l|}{D} & \multicolumn{1}{l|}{N} & \multicolumn{1}{l|}{\begin{tabular}[c]{@{}l@{}}Nterm\\ score\end{tabular}} & \multicolumn{1}{l|}{\begin{tabular}[c]{@{}l@{}}Cterm\\ score\end{tabular}} \\ \hline
		\multicolumn{1}{|l|}{$Best~Nterm_{parent}$} & \textbf{AAA}PEPSEQK & 0.1173 & 0.118 & 0.14 & 0.060 & 2 & 0 \\ \hline
		\multicolumn{1}{|l|}{$Best~Cterm_{parent}$} & PEPSEQ\textbf{AR} & 0.4477 & 0.237 & 0.014 & 0.025 & 0 & 2 \\ \hline
		\multicolumn{1}{|l|}{$Best~helper_{parent}$} & RG\textbf{LAAAD}TK & 0.2952 & 0.309 & 0.002 & 0.011 & 0 & 0 \\ \hline
		\multicolumn{1}{|l|}{$Offspring$} & \textbf{AAALAAADAR} & 2.1950 & 0.595 & 0.000003 & 0.000 & 8 & 8 \\ \hline
	\end{tabular}
	\vspace{-5mm}
\end{table}
The second example in Table \ref{table:CX_short_parents} shows two parents with only a few number of exact match amino acids. As the concatenated sequence still does not meet the mass difference criterion, the third parent is used to fill the gap. It can be seen from the fitness values of the helper parent that it is not necessary to have a high Nterm or Cterm score, as the helper parent is chosen based on its overall fitness value. Here also an exact match is obtained, but it is worth mentioning that applying this operator does not always results in whole sequence exact match, but mainly there is an improvement in the fitness value of the new offspring. 

Table \ref{table:CX_overal_performance} presents the overall performance of Nterm-Cterm Crossover on a number of individuals produced by tag-based initialisation method. In the first row, the tag-based initialisation method is used in 30 independent runs, each run producing 1000 individuals. In each run, out of 1000 individuals three individuals with having the best Nterm score, Cterm score and fitness value are chosen to be Nterm parent, Cterm parent and helper parent, respectively. Then the Nterm-Cterm Crossover is applied on the parents of each run and the average delta fitness values are calculated for all the runs. It can be seen that in overall the fitness values of the offsprings improved by 62\% compared to the Nterm parent, 37\% to Cterm parent, and 28\% compared to the helper parent which is the best individual in each run.

Similarly, the second row of Table \ref{table:CX_overal_performance} presents the results of improvement in the fitness values of new offsprings produced by Nterm-Cterm Crossover in a single run, but randomly choosing 30 individuals as parents which are not necessarily the best scored parents. The results show that in this case also in average there is 4\% improvement in the fitness score of the new offspring compared to its Nterm parent, 11\% compared to Cterm parents and 1.4\% compared to the helper parent. One reason of not having a significant improvement in this results is that the parents are not filtered. That is why in design of the GA algorithm, presented in Fig. \ref{fig:GA_Novo_workflow}, the individuals in two Nterm and Cterm pools must have at least an Nterm/Cterm scores of one.

\begin{table} [t]
	\centering
	\caption{Performance evaluation of Nterm-Cterm Crossover operator using the spectrum of \textquotedblleft AAALAAADAR\textquotedblright peptide in different scenarios.}
	\label{table:CX_overal_performance}
	\vspace{-3mm}
	\setlength{\tabcolsep}{5 pt}
	
	\begin{tabular}{c|c|c|c|}
		\cline{2-4}
		& $\Delta f_{cx,N_{parent}}$ & $\Delta f_{cx,C_{parent}}$ & $\Delta f_{cx,H_{parent}}$  \\ \hline
		\multicolumn{1}{|c|}{\begin{tabular}[c]{@{}c@{}}30 runs \textquotedblleft Best" individuals\end{tabular}} & 0.62 & 0.37 & 0.28  \\ \hline
		\multicolumn{1}{|c|}{\begin{tabular}[c]{@{}c@{}}single run \textquotedblleft Random" individuals\end{tabular}} & 0.4 & 0.11 & 0.014 \\ \hline
	\end{tabular}
	\vspace{-3mm}
\end{table}

\vspace{-4mm}
\section{Conclusions and Future Work}
\vspace{-1mm}

The goal of this paper was developing an effective \textit{de novo} sequencing algorithm that constructs full length sequences. The goal has been successfully achieved by developing an effective GA algorithm that gradually and rapidly construct the peptide sequences that match the input MS/MS spectra. 

Other developments presented in this work are a new domain dependant fitness function,  new initialisation method and two new genetic operators that were particularly designed for the GA algorithm. The GA fitness function was able to capture main spectral features and guided GA to produce the fully matched peptides. The initialisation method was an excellent start point to accelerate the evolutionary process. The tag-based initialisation method helped GA to start with better/fitter initial population, accelerating its convergence speed, and providing high quality individuals for the GA components. The genetic operators helped GA to maintain the diversity in the population and gradually convert partial matches to fully matched sequences. The results showed that GA-Novo achieved higher number of fully matched sequences compared to PEAKS, the most commonly used \textit{de nevo} sequencing software. GA-Novo achieved both higher recall and precision than PEAKS. GA-Novo outperformed PEAKS by 4\% higher precision, 4\% higher recall in amino acid level and 8\% higher recall in peptide level.

As future work, we will investigate the performance of GA-Novo using post-transnationally modifies spectra. Designing different types of mutation operators that substitute an amino acid according to a probability from a substitution matrix (for example BLOSUM62 matrix) or from a list of known PTMs will be considered in our next work. At the same time we will also work on improving the performance of the system at peptide level to get more fully matched sequences by considering a dictionary of di-peptide conflict masses.
\vspace{-4mm}

%
%

%
%
%
%
\bibliographystyle{unsrt}
\bibliography{Reference}
\end{document}